\documentstyle[11pt]{article}
\begin{document}
\thispagestyle{empty}
%\begin{flushright} December 1999\\
%\end{flushright}
\vspace{0.5in}
\begin{center}

{\Large \bf Effects of New Gravitational Interactions on Neutrinoless
Double Beta Decay}

\vspace{1.0in}

{\bf H.V. Klapdor--Kleingrothaus$^1$, H. P\"as$^1$ and U. Sarkar$^2$\\}

\vspace{0.2in}

\noindent {$^1$ \sl Max--Planck--Institut f\"ur Kernphysik,
P.O. Box 103980, D--69029 Heidelberg,
Germany \\}
\noindent{$^2$ \sl Physical Research Laboratory, Ahmedabad 380 009, India\\}

\vspace{1.0in}

\begin{abstract}

It has recently been proposed that violations of Lorentz invariance
or violations of the equivalence principle can be constrained from
the non-observation of neutrinoless double beta decay. We 
generalize this analysis to all possible new gravitational 
interactions and discuss briefly the constraints for different cases.

\end{abstract}
\end{center}

\newpage
\baselineskip 18pt

Although there is no evidence for the violation of gravitational
laws, lots of work has been done to find out to what accuracy
this is true. Many experiments have been performed to test the 
equivalence principle \cite{rel} for ordinary matter and to test
local Lorentz invariance \cite{will,relc}. In recent times there
has been some effort to test these laws for the gravitational 
couplings of neutrinos. Assuming that neutrinos of
different generations have characteristic couplings 
to gravity with differing strength and that the gravitational 
eigenstates differ from the mass eigenstates, one can constrain
the amount of violation of equivalence principle (VEP)in the 
neutrino sector from present neutrino oscillation results
\cite{vep1,vep2}. 
Similar bounds were also obtained for the amount of 
violation of local Lorentz invariance (VLI), assuming 
that neutrinos of different generations have characteristic 
maximum attainable velocities \cite{vli1,vli2}. 
Recently we have pointed 
out that in both these cases it is possible to constrain
some otherwise unconstrained region in the parameter space from 
neutrinoless double beta decay \cite{beta}. 

In this article we propose a general framework to 
study the effect of new gravitational interactions 
in the neutrinoless double beta decay. This formalism is 
similar to the one used in the study of $K-$system 
\cite{nacht,vepk}. We classify all possible
interactions as scalar, vector and tensor interactions. 
Since both the VEP and VLI are tensor interactions, 
it is expected that in both cases similar
constraints should be obtained, as observed. On the other hand, a recent
string motivated violation of the equivalence principle
a la Damour and Polyakov \cite{pd} is a scalar interaction.
Thus the constraint in this case is of different nature than in
the cases of VEP or VLI considered previously. 
The possibile fifth force \cite{fifth} discussed
in the literature is a vector interaction and thus also has a
different phenomenology. Our analysis can be 
extended to study the effects of gravitational interactions
in neutrino oscillation experiments.

We write down the most general lagrangian for
interactions of  
neutrinos with scalar, vector and tensor fields in the
weak basis $[\nu_i]$ following the general framework
developed for the $K-$system \cite{nacht},
\begin{equation}
{\cal L} = G_{ij} \nu_i \nu_j
+ G_{ij}^\mu \nu_{i,\mu} \nu_j
+ G_{ij}^{\mu \nu} \nu_{i,\mu} \nu_{j,\nu}
\end{equation}
where $i,j$ are generation indices and $ G_{ij}$,
$ G_{ij}^{\mu}$ and $ G_{ij}^{\mu \nu}$ are
scalar, vector and tensor fields respectively. We 
shall not work beyond the external field 
approximation. These fields have some restrictions
coming from the symmetry properties and by discarding
the total divergence expressions from the lagrangian,
which have been discussed in ref. \cite{nacht}
in detail. We further assumed that the gravitational
eigenstates could be different from the mass 
eigenstate as well as the weak eigenstate.
For simplicity from now on we shall work in an
two generation scenario, $i,j = e, x$ with $x=\mu,\tau,s$. 

We now can write down the Feynmann 
diagrams and hence the self energy matrix in the same way
as in ref \cite{nacht}, from which 
the contribution to the effective hamiltonian can be read off 
in the weak basis, given by
\begin{equation}
{\cal H}_{ij} = G_{ij} +
i G_{ij}^{\mu} p_\mu + G_{ij}^{\mu \nu} p_{\mu} p_{\nu}
\label{pmu} .
\end{equation}
This hamiltonian is related to the effective hamiltonian in the mass
and gravitational bases through unitary rotations
\begin{equation}
H = U_{m} H_{m} U_{m}^{-1} + U_G H_G U_G^{-1} \label{h}.
\end{equation}
In absence of any new gravitational interactions 
the neutrino mass matrix in the mass basis 
$[\nu_1 ~~ \nu_2 ]$ is given by
\begin{equation} 
H_{m} = \frac{(M_{m})^2}{2 p} = \frac{1}{2 p} {\pmatrix{
m_1 & 0 \cr 0 & m_2 }}^2 \label{hsew}
\end{equation}
and the gravitational interaction part of the hamiltonian is
\begin{equation}
H_G = pI + \frac{(M_{G})^2}{2 p} = p I + \frac{1}{2 p} \pmatrix{ 
g_1^a & 0 \cr 0 &  g_2^a}  .
\end{equation} 
Here $p$ denotes the momentum, $I$ represents an unit matrix and 
$\bar{m}$ the average mass, and for any quantity $X$ we define $\delta X
\equiv(X_1-X_2)$, $\bar{X} = (X_1+X_2)/2$. $a=S,V,T$ represents
scalar, vector and tensor interactions respectively.

The scalar, vector and tensor gravitational interactions
can be written in the following forms so as to reproduce
the correct dimensions of equation (\ref{pmu}),
\begin{eqnarray}
g_i^S &=& 2 {{\alpha^S}_{i} m_i}^2  \nonumber \\
g_i^V &=& 2 {\alpha^V}_i m_i  p \nonumber \\
g_i^T &=& 2 {\alpha^T}_i p^2 \nonumber
\end{eqnarray}
In the absence of any gravitational interactions ${\alpha^a}_i = 0$,
$H_G$ simply becomes the momentum of the neutrinos. Here
we are interested in a single virtual neutrino 
propagating inside the nucleus with a particular momentum. For this 
reason we assume the momenta of both the neutrinos are $p$ in the
absence of any new gravitational interactions. 
Hence $\alpha^a_1 - \alpha^a_2 = \delta \alpha^a $ is a measure 
of  the new gravitational interactions in the 
neutrino sector. To compare our result with the neutrino
oscillation experiments we further assume,
$\alpha^a_1 + \alpha^a_2 = 0$, {\it i.e.}, there is no mean 
deviation from the gravitational laws and there is only a
relative violation given by the measure $\delta \alpha^a$.
This approximation will reduce the the number of parameters
so that we can compare the bounds from the neutrino 
oscillation experiments with the ones from neutrinoless 
double beta decay. 

We tried to keep the above discussions as general as possible
with the restrictions that we donot go beyond the perturbative
regime. We assume that the corrections to gravity comes from
interactions with some external scalar, vector or tensor 
fields only and there is no non-renormalizable higher 
dimensional operators which modifies gravity with inverse mass
scales. Our general parametrization has one drawback that 
although we are working in the gravitational basis, the masses
involved in the expressions for $g_i$ are considered in the 
mass basis. This can be justified by assuming VEP to be a small effect.
In the case of tensor interaction masses donot enter, only in
the scalar and vector cases this problem appears. However, as
we shall see, the final result for the scalar case comes
out to be the same as the one derived from other approaches \cite{dil}.
Moreover, in the case of some scalar interactions the gravitational
basis is equal to the mass basis \footnote{this point will
be discussed in a forthcoming article, where the dilaton-exchange
gravity will be studied by the authors}, and then this question 
will not arise. 

We shall not consider any $CP$ violation, and hence $H_{m}$ and
$H_G$ are real symmetric matrices and $U_m$ and $U_G$ are orthogonal
matrices $U^{-1}= U^T$. They can be parametrized as 
$U_i = \pmatrix{\cos \theta_i & \sin \theta_i \cr -\sin \theta_i
& \cos \theta_i}$, where $\theta_i$ represents weak mixing angle $\theta_m$ 
or gravitational mixing angle $\theta_G$. We can now write down the 
weak Hamiltonian $H_w$ in the weak basis, 
in which the charged lepton mass matrix is diagonal
and the charged current interaction is also diagonal, as
\begin{equation} 
H = p I +  \frac{1}{2 p} { \pmatrix{ M_{+} & M_{12} \cr
M_{12} & M_{-} }}^2 = p I +  \frac{1}{2 p}
\left( M_1 + {1 \over 2} M_2^2 M_1^{-1} \right)^2 .
\end{equation}
where, 
\begin{eqnarray}
M_1 &=& U_m M_m U_m^{-1} \nonumber \\
M_2 &=& U_G M_G U_G^{-1} \nonumber
\end{eqnarray}
and we assumed $M_1^2 \gg M_2^2$, so that the gravitational effects
are much smaller than the usual neutrino masses. Since no new
gravitational effects have been observed so far, we use this formalism
to constrain the parameters of the new gravitational interactions, for
which this assumption is justified. We then obtain,
\begin{eqnarray}
M_{\pm} &=& \bar{m} \pm \frac{cos2\theta_m}{2}
\delta m \nonumber  \\
&&\pm \left[ \mp \bar{g^a} \bar{m} - \delta g^a \bar{m} \frac
{\cos2\theta_G}{2} \pm {\delta m \delta g^a \over 4} 
\cos 2 (\theta_G - \theta_m) + \delta m \bar{g^a} {\cos 2 \theta_m 
\over 2} \right] / [2 ( \delta m^2 - \bar{m}^2)] \nonumber \\
M_{12} &=& -{\sin2\theta_m \over 2} \delta m \nonumber \\
&& + \left[ \delta g^a \bar{m} {\sin2\theta_G \over 2} 
- \delta m \bar{g^a} {\sin 2 \theta_m \over 2} \right] / 2 ( 
\delta m^2 - \bar{m}^2)  .
\end{eqnarray}
to a leading order in $\delta g^a $. 

The decay rate for the neutrinoless double beta decay is given by,
\begin{equation} 
[T_{1/2}^{0\nu\beta\beta}]^{-1}=\frac{M_+^2}{m_e^2} 
G_{01} |ME|^2,
\end{equation} 
where $ME$ denotes the nuclear matrix element, $G_{01}$
corresponds to the phase space factor defined in \cite{doi} and $m_e$ is the 
electron mass. The momentum dependence of $M_+$ must be absorbed 
into the nuclear matrix element, so that this quantity
contains all the momentum dependence and the remaining part is 
estimated using zero momentum transfer approximation. 
Thus, if one ignores the nuclear matrix element, then obviously 
there cannot be any effect of the vector and tensor type 
gravitational interactions in
neutrinoless double beta decay, which was mistaken in ref. \cite{wrong}.
As it has been discussed earlier \cite{beta}, the momentum dependence
of the tensor type gravitational interactions enters the nuclear 
matrix element, which then is enhanced by a factor $p^2$ coming
through $M_+$ in the above expression. 

We shall now present a more
detail explanation of this analysis.

In ref \cite{wrong} it is claimed that neither violations of
Lorentz invariance nor violations of the equivalence principle may
give sizable contributions to neutrinoless double beta decay. 
The argument discussed is the following: Taking the neutrino
propagator 
\begin{equation} 
\int d^4 q \frac{e^{-i q (x-y)} \langle m \rangle c_a^2}{m^2 c_a^4 
- q_0^2 c_a^2 + \vec{q}^2 c_a^2 }
\end{equation} 
with the standard $0\nu\beta\beta$ observable $\langle m \rangle$, the
neutrino four momentum $q$ and the characteristic maximal velocity $c_a$.
If one would neglect now $q_0$ and $m$ in the denominator, $c_a$ drops
out and the decay rate is independent of $c_a$.

However, in \cite{beta} it has been shown starting from the Hamiltonian level
that the propagator (or the $0\nu\beta\beta$ observable) is changed 
itself violating Lorentz invariance. Since
\begin{eqnarray} 
H&=&\vec{q} c_a + \frac{m^2 c_a^4}{2 \vec{q} c_a} \nonumber  \\
&=& \vec{q} I + \frac {m^{(*)2} c_a^4}{2 \vec{q} c_a} \label{1}
\end{eqnarray} 
with $c_a = I + \delta v$ and $m^{(*)2}=m^2 + 2 \vec{q}^2 c_a \delta v $
an additional contribution is obtained $\propto \vec{q}^2 \delta v$.
This mass-like term has a $\vec{q^2}$ enhancement and is not
proportional to the small neutrino mass.
This consideration answers also the frequently asked question ``What
is the source of lepton number violation?'' in this mechanism.
Comparable to a usual mass term, which can be both of Majoran type as well
as Dirac type the mass-like term 
$2 \vec{q}^2 c_a\delta v $ can be of Majorana type and act as the source
of lepton number violation in this context.

We shall now discuss the three different cases of scalar, vector
and tensor interactions and their phenomenology. In the case of 
tensor interaction the constraint has already been discussed in
ref.  \cite{beta}. The violation of local Lorentz invariance and
the violation of the equivalence principle both fall under this
category (their equivalence has been pointed out elsewhere
\cite{will}). In both these cases the effect of the new 
gravitational interactions have quadratic momentum dependence.

In case of the tensorial gravitational interaction we have,
$\bar{g^T} = 2 \bar{\alpha^T} p^2 = 0$ and $\delta g^T
= 2 \delta \alpha^T {p^2}$. 
In particular, for the violation of the
equivalence principle we substitute $\delta \alpha^T = 
4 \delta g \phi $ (following the notation of ref \cite{beta}),
where $\phi$ is the Newtonian gravitational potential on the
surface of earth. On the other hand, for the violation of
the local Lorentz invariance, we substitute instead,
$\delta \alpha^T = 2 \delta v $. In both these cases bounds 
were given in ref \cite{beta}. 

To give a bound on tensorial gravitational interactions 
in the small mixing region  (including $\theta_v \sim \theta_m \sim 0$)
conservatively $\langle m \rangle \simeq 0$ was assumed.
It was also assumed that
$\delta{m} \leq \bar{m}$, and thus $\frac{\delta m}{4 \bar{m}}$ 
may be neglected. 
Due to the $p^2$ enhancement the nuclear matrix elements of the 
mass mechanism have to be replaced by $\frac{m_p}{R}\cdot 
(M_F^{'}-M_{GT}^{'})$ 
with the nuclear radius $R$ and the proton mass $m_p$, which have been 
calculated in \cite{mat}.
Inserting the recent half life limit obtained from the Heidelberg--Moscow 
experiment \cite{double},  
a bound on the amount of tensorial gravitational interactions 
as a function of the average neutrino mass
$\bar{m}$ was given \cite{beta}. It should be stressed also 
that the GENIUS proposal of the Heidelberg group
\cite{gen} could improve these bounds  
by about 1--2 orders of magnitude. 

For the vector type gravitational interactions there is 
a linear momentum dependence. In this case,
$\bar{g^V} =2 \bar{\alpha^V} p m = 0$ and $\delta g^V
= 2 \delta \alpha^V p m$. The fifth force, as
discussed by Fishbach et al \cite{fifth} in the context of 
$K-$physics is a vector type gravitational interaction. Since
no studies of this type of forces exist for neutrino 
oscillation experiments, with which neutrinoless double beta decay 
results could be compared, we shall not study this case.

A similar generic structure was considered in a recent
analysis of the atmospheric neutrino anomaly \cite{lisi},
where they used the power of momentum dependence
as a parameter. 
From their analysis it becomes apparent that 
the atmospheric neutrino 
anomaly may not be explained by either tensorial or 
vectorial gravitational analysis alone \cite{lisi}.

Recently it has been argued by Damour and Polyakov \cite{pd}
that string theory may lead to a new scalar type 
gravitational interaction through interaction of the dilaton
field and subsequently its consequence to neutrino 
oscillation has been studied \cite{dil}.
Damour and Polyakov have shown that the massless dilaton 
interaction modifies the gravitational potential energy and
there is an additional contribution from an spin-0 exchange,
which results in a scalar type gravitational interaction \cite{pd}. 
The resulting theory is of scalar-tensor type with the two particle
static gravitational energy
\begin{equation} 
V(r)=- G_N m_A m_B (1+\alpha _A \alpha_B)/r,
\end{equation} 
where $G_N$ is Newton's gravitational constant and $\alpha_j$ denotes the
couplings of the dilaton field $\phi$ 
to the matter field $\psi_j$, leading to a gravitational energy of
\begin{equation} 
L= m_j \alpha_j \overline{\psi_j}\psi_j \phi.
\end{equation} 
Thus the modified effective mass matrix of the neutrinos are now
given by \cite{dil}
\begin{equation} 
m^{(*)}=m-m \alpha \phi_c
\end{equation} 
where, the classical value of the dilaton field $\phi_c=\phi_N 
\alpha_{ext}$ is characterised by the $\alpha$ value of the bulk 
matter producing it and for a static matter distribution 
proportional to the Newtonian potential $\phi_N$. 

The effective mass squared difference 
\begin{equation} 
\Delta m^{(*)2}=-2 m^2 \phi_N \alpha_{ext} \delta \alpha
\end{equation} 
(for almost degenerate masses $m$) gives rise to neutrino oscillation.
The corresponding effect for $0\nu\beta\beta$ decay is
obtained by replacing $\delta g^S = 2 \delta \alpha^S m^2$ (for
almost degenerate mass $m_1 \sim m_2 \sim m$).
Comparing the arguments in the oscillations propabilities we get
\begin{equation} 
M_+=m + m \alpha_{ext} \Phi_N \delta \alpha 
\frac{\cos(2 \theta_G)}{2}  .
\end{equation} 

In this case, it is difficult to obtain any bound from neutrino
experiments since for $\alpha_{ext}$ only upper bounds exist. 
However, to get an idea of the constraints which can come from
neutrino experiments in the future if $\alpha_{ext}$ is known, 
according to 
ref. \cite{dil} we assume $\phi_N=3 \cdot 10^{-5}$,
$\alpha_{ext}=\sqrt{10^{-3}}$ and $m=2.5$ eV (as an upper 
bound obtained from tritium beta decay experiments \cite{tritium}).
In this case the quantity $\delta \alpha$ is not constrained from
neutrinoless double beta decay. 

 In summary, we presented a general formalism for the study of 
effects of new gravitational interactions in 
neutrinoless double beta decay, which 
allows to constrain the amount of violation of the 
gravitational laws. Various scenarios
discussed in the literature have been analyzed as special cases
of the present formalism. 

\section*{Acknowledgement}
We thank G. Bhattacharyya for useful discussions.

\end{document}